\documentstyle[prd,aps]{revtex}

\def\RE {I\kern-6pt R    }
\def\Z  {Z\kern-13pt Z   }
\def\be {\begin{equation}}
\def\ee {\end{equation}  }
\def\beq{\begin{eqnarray}}
\def\eeq{\end{eqnarray}  }

\def\bi {\begin{itemize} }

\def\ei {\end{itemize}   }

\def\gtwid{\mathrel{\raise.3ex\hbox{$>$\kern-.75em\lower1ex\hbox{$\sim$}}}}
\def\ltwid{\mathrel{\raise.3ex\hbox{$<$\kern-.75em\lower1ex\hbox{$\sim$}}}}

\input epsf

\begin{document}


\title{New Critical Behavior in Einstein-Yang-Mills Collapse}


\author{Matthew W. Choptuik\footnote
        {Electronic address:  \tt matt@einstein.ph.utexas.edu\hfil}}
\address{Center for Relativity, The University of Texas at Austin, 
          Austin, TX 78712-1081\\
         Institute for Theoretical Physics, University of California,
         Santa Barbara, CA 93106-4030}
\author{Eric W. Hirschmann\footnote
        {Electronic address:  \tt ehirsch@bach.liu.edu\hfil}}
\address{Southampton College, Long Island University, Southampton, NY 11968}
\author{Robert L. Marsa\footnote
        {Electronic address:  \tt marsa@einstein.ph.utexas.edu\hfil}}
\address{Reliant Data Systems, 13915 Burnet Road, Suite 200, Austin, TX 78728}

\maketitle

\begin{abstract}

We extend the investigation of the gravitational collapse of a spherically
symmetric Yang-Mills field in Einstein gravity and show that,
within the black hole regime, a new kind of critical behavior
arises which separates black holes formed via Type~I collapse 
from black holes formed through Type~II collapse.
Further, we provide evidence that these new attracting
critical solutions are in fact
the previously discovered colored black holes with a single
unstable mode.

\end{abstract}



\section{Introduction}
\label{sec:introduction}

It is by now relatively well known that gravitational collapse can
produce rich structure even within highly simplified systems such as
spherical symmetry.  In
particular, near the threshold of black hole formation, the strong
field dynamics of general relativity exhibits
critical phenomena.

The pioneering work demonstrating this critical behavior in
the collapse of a single massless scalar field \cite{Matt} has been
supplemented by investigations of
gravitational waves \cite{AbeEvns}, a perfect fluid \cite{EvnsCol} and a
variety of other matter models, all of which exhibit the same
general characteristics.
Indeed, to our knowledge no system which has been studied in this
context has been shown {\em not} to exhibit this critical phenomena.

At this point in the subject's development,
dynamical evolutions (solutions of the full partial differential 
equations of motion), in tandem with analytic
and perturbative calculations have given us a reasonable understanding
of many of the phenomenological details of critical behavior in collapse.
(See~\cite{Gundlach} for an excellent review of the subject.)

In light of this, one of the more interesting discoveries in some of the
more recently studied
models~\cite{ChopBizChmj,BradChamGonc,BizonChmaj98,BizonChmaj98a,BCT}
is the presence of two distinct types of behavior at the threshold
of black hole formation.
Specifically, in these models, certain regions of parameter space
(initial-data space) are found to yield near-critical collapsing
configurations which display self-similarity and, in the
super-critical regime, a scaling law which is continuous in the
black hole mass.
By analogy with the theory of phase transitions, this is called a Type
II transition.  However, it is found that other regions of parameter space
lead to critical collapse which has a static (or periodic)
solution as an intermediate attractor---this results in a black
hole transition with a nonzero mass gap.  Again, in analogy with the
discontinuous behavior in order parameter which frequently accompanies
first order phase transitions, this is called Type~I behavior.
For the case of the Einstein-Yang-Mills (EYM) model studied
in~\cite{ChopBizChmj}, the static solution appearing in Type
I collapse is the well-known $n=1$ Bartnik-Mckinnon
solution\cite{BMK}, while for the
massive scalar model considered in~\cite{BradChamGonc}, the Type
I threshold solutions are apparently unstable members of the
family of ``oscillating soliton stars'' which have previously been
constructed by Seidel and Suen~\cite{SeidelSuen91}, albeit in
a different context.
Finally, in the case of Einstein-Skyrme (ES) collapse considered
in~\cite{BizonChmaj98,BizonChmaj98a}, a static Type~I solution was
observed, which, as in the EYM case, had previously been constructed
and studied within a purely static {\em ansatz}~\cite{BizonChmaj92}.
As noted above, each of these three models also exhibits Type~II
behavior---in the case of massive scalar collapse~\cite{BradChamGonc},
the Type~II critical solution is the same one originally observed
in massless scalar collapse~\cite{Matt}, and, interestingly, in the
ES model, the Type~II solution is evidently identical~\cite{BCT}
to that observed in the EYM model~\cite{ChopBizChmj}.   Heuristically,
one expects Type~II behavior in {\em any} collapse model where the
initial configuration can be made sufficiently ``kinetic-energy-dominated''
(ultra-relativistic), whereas Type~I behavior is expected only in those
models which have an intrinsic length scale (or equivalently, mass scale),
and which have some type of self-interaction which can ``balance'' the
attractive gravitational interaction.

As noted in the concluding remarks of~\cite{ChopBizChmj}, hints of
further interesting phenomenology in the EYM model have been seen
in the super-critical regime where all evolutions are characterized
by black hole formation.  In this paper we study this regime in
more detail and present evidence for a new type of critical transition
in which the intermediate attractors are the ``minimally unstable''
(one unstable mode in perturbation theory) colored black holes
discovered by Bizon \cite{Bizon}, and independently by Volkov and
Gal'tsov \cite{VolGal}. This result is, of course, analogous to the
discovery that the $n=1$ Bartnik-Mckinnon solution is the intermediate
attractor for Type~I collapse.  Crucially, in order to accurately
model the dynamics of super-critical solutions for long times after
the formation of an event horizon, we use black-hole excision techniques.
Such methods were first successfully employed in a dynamical context by
Seidel and Suen~\cite{SeidelSuen92}, 
and have subsequently been studied and implemented by
many other authors (see~\cite{MarsChop} and references therein for
a more extensive discussion).  However, to our knowledge, this is the
first time that excision has been used to study critical collapse, and
we feel our results highlight the power and potential of the strategy to
elucidate issues relating to the formation and long-time 
evolution of black holes.
Our adoption of excising techniques necessitates the use of a
different coordinate system than that used in~\cite{ChopBizChmj}---that
work used polar slicing and areal spatial coordinates, a system which
generalizes the usual Schwarzschild coordinates to time-dependent
spacetimes.  As is well-known, the $t=\,$constant slices in the
polar/areal system cannot penetrate apparent horizons---thus, for
all practical purposes, the slices remain outside of event horizons
and therefore cannot be used in conjunction with excision.  We therefore
retain areal spatial coordinates, but adopt maximal slicing---in this
case the slices {\em do} cross apparent and event horizons, and
excision techniques {\em can} be used.

The outline of the remainder of the paper is as follows.  The next
section
describes the EYM model and the equations of motion we
subsequently solve numerically.
We pay particular attention to gauge and coordinate choices, as well
as to regularity, boundary and initial conditions for the fields.
In Section~\ref{sec:excision} we describe our numerical scheme, focusing
on our specific coordinate choices and on some details of our
black-hole excising technique.  In Sections~\ref{sec:results} 
and~\ref{sec:discussion}
we describe our results and conclusions.

\section{Equations and Assumptions}
\label{sec:equations}
We are interested in investigating the gravitational
collapse of a self-gravitating
Yang-Mills gauge field.
To begin, let us consider the action for an EYM theory
\be
S = \int d^4x \sqrt{-g} \left[{R \over 16\pi G}
     - {1\over g^2} F^{a}_{\mu\nu} F^{a\mu\nu}
     \right]
\ee
where $F^{a}_{\mu\nu}$ is the Yang-Mills field strength tensor.
On varying the action with respect to the metric, $g_{\mu\nu}$,
and the gauge connection, $A^{a}_{\mu}$, we get the general equations of
motion.  We simplify these further by making some additional assumptions.
In particular, we choose the gauge group to be $SU(2)$ and focus 
on spherically symmetric gravitational collapse.  This places restrictions
on both the spacetime metric and the form of the gauge connection.
Even so, the equations we derive have a rather general form as can be seen
in Equations (\ref{eom_first}--\ref{eom_last}).  Subsets of these equations
have been considered in a variety of different contexts.
In particular
\cite{ChopBizChmj} evolved a version of the Einstein-Yang-Mills equations
in polar
($K^{\theta}{}_{\theta} = 0$), areal ($b=1$) coordinates with the additional
assumption on the Yang-Mills field that the connection $A_\mu{}^a$ was purely
magnetic.

Since our interest here is to consider the same model 
studied in~\cite{ChopBizChmj}, 
but to penetrate into
the super-critical regime of the phase space, we will likewise make
the ``magnetic {\em ansatz}."  As described in the Appendix, together with
appropriate gauge choices, this {\em ansatz} effectively sets
all but one of the components of the gauge connection to zero.

In addition to making these gauge choices we must also
choose a coordinate system.
As mentioned in the Introduction, in order to evolve the system for long
times to the future of black hole formation, we choose maximal time
slices and areal (or radial) spatial coordinates. 

As detailed in the Appendix, the equations can now be written
in the following form.  The evolution equations are
\beq
\label{eq:evpi}
\dot{\Pi} & = & \left[ \beta\Pi + {\alpha \over a} \Phi \right]'
     + {\alpha a\over r^2} w (1-w^2)  \\
\dot{\Phi} & = & \left[ {\alpha \over a} \Pi + \beta\Phi \right]'
          \\
\label{eq:evw}
\dot{w}         & = & {\alpha \over a} \Pi + \beta w'  
\eeq
and the constraint equations are
\beq
w' & = & \Phi  \\
\label{eq:alpha}
\alpha'' & = &  \alpha' \left( {a'\over a} - {2\over r} \right)
         + {2\alpha\over r^2} \left( a^2 - 1 + {2r a'\over a} \right)
 + 4\pi G \alpha \left( S - 3 \rho \right)   \\
\label{eq:a}
a' & = & a {1-a^2 \over 2r} + {3\over2} ra^3 K_{\theta}{}^{\theta}{}^2
 + 4\pi G r a \rho   \\
\label{eq:Ktt}
K_{\theta}{}^{\theta}{}' & = & -{3\over r} K_{\theta}{}^{\theta}
    + 4\pi G \left( {\Pi\Phi\over g^2 a r^2}  \right)
\eeq
where the matter stress-energy terms are given by
\beq
S - 3 \rho & \, = & \,
      { a^2 (1-w^2)^2 \over 2 g^2 r^4} + {1 \over g^2 r^2 }
         \left( \Phi^2 + \Pi^2 \right)      \\
\rho & \, = & \,
      { a^2 (1-w^2)^2 \over 4 g^2 r^4} + {1 \over 2 g^2 r^2 }
         \left( \Phi^2 + \Pi^2 \right) .
\eeq
We also note that we have an algebraic relation for the sole component, 
$\beta$, of the shift vector, $\beta^i = (\beta,0,0)$:
\beq
\beta = \alpha r K_{\theta}{}^{\theta} \, .
\eeq

In addition to the equations of motion, we need boundary
conditions on the
fields.  These are determined by demanding regularity
at the origin of spherical symmetry, and by enforcing an outgoing
condition on the radiation fields at large radius.  This latter condition
assumes that there is no radiation coming in from outside our finite mesh.
This is not completely true, as in general there will be backscattering of the
propagating fields off regions of high curvature.  However, if our domain of
integration is large enough, the contributions from this scattering are
dynamically negligible.

The demand for a regular origin results in boundary
conditions
on the fields at $r=0$.  From 
equations~(\ref{eq:alpha}) and (\ref{eq:a}) it can be seen
that $\alpha'(t,0) = a'(t,0) = 0$, as well as
$a(t,0) = 1$.  In addition, $\beta(t,0) = K_{\theta}{}^{\theta}(t,0) = 0$. 
For the matter fields, $w(t,0)=\pm1$ and $w'(t,0)=0$, so that the auxiliary 
variables $\Pi$ and $\Phi$ are both zero at the origin.  We note that these
conditions constrain the Yang-Mills field, $w$, to be in a vaccum state 
at $r=0$.

For initial data, we choose a time-symmetric
kink for the gauge potential which
was previously used in \cite{ChopBizChmj}.
This pulse is given by
\beq
\label{eq:kink}
w(0,r) & \, = & \, \left[ 1 + a \left(1 + {br \over s}\right)
e^{-2\left(r/s\right)^2} \right]
                  \cdot \tanh \left( {x-r \over s} \right)  \\
\dot{w}(0,r) & \, = & \, 0 .
\eeq
with the parameters $a$ and $b$ chosen such that $w(0,0)=1$ and $w'(0,0)=0$. 
The two parameters $x$ and $s$ define the center and width of the pulse
respectively.
They also serve as the two parameters which we will vary in order to explore
the phase space.  We note that the implementation of this data in
\cite{ChopBizChmj} was incorrect but ultimately
had no effect on the overall conclusions
of that work.  We have fixed the implementation of the kink data here and
note a minor improvement in the conservation of energy.

\section{Numerical approach and black hole excision}
\label{sec:excision}

It should be emphasized that we do not incorporate any form of adaptive
mesh refinement into our numerical approach.  Instead, we use a 
fixed uniform grid with a mesh spacing which is 
sufficiently fine to uncover the new critical behavior.  The drawback to
this approach, 
of course, is that we are unable to fully resolve the discretely 
self similar solutions which arise near the Type~II black hole threshold.
However, our primary interest here is in the
supercritical regime, and we are satisfied that previous work has
established the nature of the Type~II transition.

What {\em is} important in this work is the use of black hole excision
techniques which allow us to evolve well beyond the formation of 
the black hole.
We discretize the evolution equations (\ref{eq:evpi}--\ref{eq:evw}) using
two time levels with centered time differences, and angled spatial
differences as described in \cite{MarsChop}.  The constraint equations are
then integrated outward from the origin.  Equation (\ref{eq:alpha}), the
slicing equation, is
rewritten in first order form
\beq
\delta' & = &  \delta \left( {a'\over a} - {2\over r} \right)
         + {2\alpha\over r^2} \left( a^2 - 1 + {2r a'\over a} \right)
 + 4\pi G \alpha \left( S - 3 \rho \right)   \\
\alpha' &=& \delta.
\eeq

The apparent horizon equation in maximal-areal coordinates is simply
\beq
a r K^\theta{}_\theta = 1.
\eeq
Thus, we can compute a characteristic function $C$ such that
\begin{displaymath}
C  = \left\{ \begin{array}{ll}
                1&  \textnormal{if $a r K^\theta{}_\theta \leq \mu_H$},\\
                0&  \textnormal{if $a r K^\theta{}_\theta > \mu_H$}.
             \end{array} \right.
\end{displaymath}
where $\mu_H$ is a threshold value, ostensibly unity, but set to 
a slightly larger value in practice---typically, $\mu_H = 1.1$---to ensure 
that the inner boundary lies strictly within the apparent horizon.
At any time, the characteristic function tells us which grid points 
are deemed to be inside ($C(r,t)=0$) or outside ($C(r,t)=1$) the 
horizon.
As the system evolves, we recompute $C$ at
each time step and simply ``discard'' any grid points for 
which $C=0$.  The radial
locations of these grid points will then generally satisfy $r \le r_H(t)$,
where $r_H(t)$ is the instantaneous inner boundary of the computational
domain, and roughly coincides with the location of the apparent horizon.
We need {\em no}
boundary conditions for the functions evolved with equations
(\ref{eq:evpi}--\ref{eq:evw}) since the characteristic directions at 
$r = r_H$ are such that only events with $r \ge r_H$ are in the 
past domain of dependence.  We can therefore use the discrete form
of the evolution equations up to {\em and including} the inner boundary 
point.
However, the constraint equations (\ref{eq:alpha}--\ref{eq:Ktt}) must have
an inner boundary condition.  Until a horizon is formed, we can simply use
conditions derived from regularity.  Once a horizon forms, however, we need an
alternative.  If the horizon initially forms at time $t=T$ and radial 
coordinate
$r=R$, then for a function $f \in \left\{ a , K^\theta{}_\theta{}
\right\}$, we use $f(R,T)$ as the initial boundary value for $f$.  We then
evolve $f$ along this boundary using the evolution equation for $f$.  The
evolution equations that we use are as follows:

\beq
\dot{a} &=& \beta \, \left[ 4 \pi r a^3 \rho + \frac{a} {2 r} \left (1-a^2
\right) +
\frac{3}{2} \, r a^3 {K^\theta{}_\theta}^2 \right] +
r a K^\theta{}_\theta{} \, \delta - 4 \pi r a \alpha j_r \, , \\
\dot{K^\theta{}_\theta{}} &=& 3 K^\theta{}_\theta \, \left( {1\over2} \alpha K^\theta{}_\theta - {\beta\over r} \right)  
+ 4 \pi \left ( \alpha S^r{}_r - \beta j_r \right) +
\frac{\alpha}{2 r^2 a^2} \left(a^2-1 \right) -
\frac{\delta}{r a^2} \, .
\eeq
Since the lapse $\alpha$ does not have an evolution equation, we simply leave
$\alpha$ and $\delta$ ``frozen'' at whatever specific values they have when 
excision at that particular radius, $r_H(t)$, begins. 

Finally, we note that the programs which we used to generate the results that
we will describe below were written in RNPL (Rapid Numerical 
Prototyping Language) \cite{Marsa}.  This language has been specifically 
designed to 
facilitate the differencing and subsequent numerical solution of partial 
differential equations.

\section{Results}
\label{sec:results}

On evolution of the equations, we verify previously examined aspects of
this system as discussed in \cite{ChopBizChmj}.  In particular,
we find that in different regions of the parameter space, there
are both types of critical behavior at the threshold of black hole
formation.  We observe Type~I behavior, wherein black holes turn on at
finite mass and the intermediate attractor
is the $n=1$ Bartnik-Mckinnon solution.  We also confirm aspects of 
Type~II collapse in the appropriate region of parameter space.  This includes
such things as echoing
and the scaling relation for the black hole mass.  However, because
our code does not employ mesh-refinement, we are unable to fully reproduce the
results of \cite{ChopBizChmj}.
However, as discussed previously, our main emphasis here
is to address the outstanding question of what happens in the super-critical
region and to the future of black hole formation.

On evolving into the supercritical (black hole) regime, we find that
the dynamics of the resulting spacetimes exhibit two distinct
types of behavior which we describe below.  It turns out that these can be
characterized by the asymptotic vacuum state
of the Yang-Mills potential $w$.  These two behaviors are
related to the Type~I and Type~II critical collapses identified previously.

The original Type~I collapse
is described, in part, by a black hole which forms
with finite mass.  Slightly super-critical Type~I evolutions, now continued 
well to the future of black hole formation confirm this behavior, as one would
expect.  In addition, after the black hole is formed,
part of the remaining
Yang-Mills field outside the horizon radiates to infinity and some of it
falls down the black hole causing the black hole to grow. 
How much larger the initial black hole becomes through the infall of additional
matter depends, in part, on where in the parameter space the evolution
begins.
Eventually, however, and in accordance with what might be expected from
the no-hair theorems, the exterior spacetime settles down to a 
Schwarzschild black
hole with the Yang-Mills field in its vacuum state.
In general, for slightly super-critical collapse (either Type~I or Type~II),
as the remaining mass-energy radiates away the gauge potential approaches
one of its two vacuum values:  $w=\pm1$.  For the case of slightly
super-critical Type~I collapse,
in the asymptotic regime, the gauge potential always picks the $w=-1$
vacuum state (recall that our initial data family~(\ref{eq:kink}) has 
$w(t,0) = +1$ and $w(t,\infty) = -1$).
This turns out to be true even for collapsing configurations
which are no longer just slightly super-critical but are, in fact, well
into the black hole regime.  We can thus characterize
a sort of ``generalized" super-critical Type~I collapse by
the asymptotic value of the gauge potential $w$.

When Type~II collapse is considered, the overall picture is broadly the
same.  Again, after the
original black hole forms (at {\em finite} mass
since we
are slightly super-critical) some of the matter remaining in the region
exterior to
the black hole falls into the black hole causing it to grow and
the rest is radiated off to infinity.  The resulting exterior spacetime 
agains settle down to Schwarzschild with the Yang-Mills field in its vacuum
state.  However, the gauge potential in the slightly super-critical Type~II
collapse approaches
its other vacuum value:
$w\rightarrow+1$.  Similarly, further into the Type~II super-critical regime
(where we should emphasize
the discretely self-similar
nature of the critical solution is no longer clear and where
the black holes which form all have finite mass) we find ``generalized"
super-critical Type~II
collapse which can be characterized by an asymptotic value of the
gauge potential of $w=+1$.

It is worth commenting here that our terminology has become a bit loose.
We have been speaking of Type~I and Type~II collapse well into the
super-critical regime
when in fact these terms originally referred
to types of critical behavior
in a very small neighborhood of the exact critical solutions.
For instance, as one moves away from the Type~II threshold, the discrete
self-similarity of the exactly critical solution is no longer clearly exhibited
by configurations which collapse to a black hole.
In addition, a significant fraction of
the mass present will end up in the black hole.  This is to be contrasted
with the tiny
black holes that can be formed in near-critical Type~II collapse.
It should also be made clear that Type~I and Type~II do not refer to
different {\em types} of black holes.  The end-state exterior geometry in all 
of these evolutions is a Schwarzschild black hole.  
However, we are, in a sense, generalizing
the concepts of Type~I and Type~II collapse to distinguish two kinds of
collapse {\em dynamics} in the super-critical regime.

\begin{figure}
\epsfxsize=15cm
\centerline{\epsffile{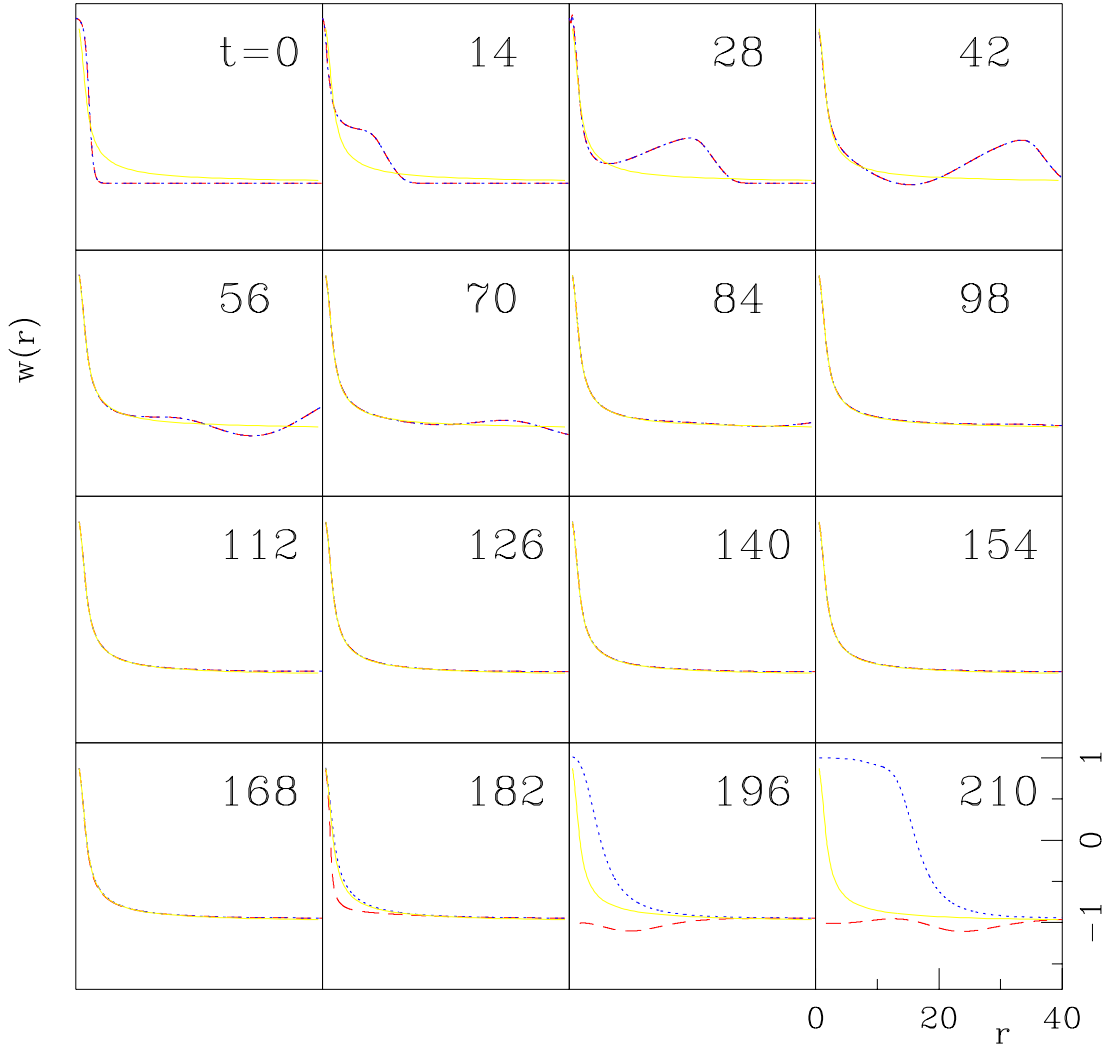}}
\caption[Snapshots of near critical evolution.]
{Time sequences of the Yang-Mills potential $w$ for
both sub- and super-critical evolutions, overlayed with a static, colored
black hole solution.  Both evolutions are calculated to machine precision,
{\it i.e.} $|p-p^{*}|/p^{*} \ltwid 10^{-15}$.  The solid line is the static
solution, the dashed line is the generalized Type~I solution, where part
of the Yang-Mills field of the colored black hole collapses
to a second, larger black hole, and the dotted line is the generalized
Type~II solution, in which the remaining Yang-Mills field surrounding the
colored black hole disperses to infinity.  The calculations shown here
use the kink data defined in Eq.(\ref{eq:kink}).
We fix the center of the pulse to be
$x=2.16$, and perform our parameter
search on $s\,(=\,p)$, the width of the pulse.  These plots use
the original radial coordinate, $r$.  The initial colored black hole that
forms has a
radius of $r\approx0.55$ on a domain from $r=0$ to $r=40$; a fixed uniform 
radial grid with 3201 points is used for all calculations.
In the generalized Type~I case, the
second and larger black hole forms with a radius of $r\approx1.69$.  Note 
that
after the black hole forms, we do not plot points at locations within
the horizon radius.  This illustrates that our horizon excision
continues the evolution only outside the black hole.  As can be seen, the
agreement between the static solution and the intermediate attractor to
which the dynamical solutions temporarily evolve is excellent.  
The static solution is
obtained by solving the static equations (ODEs) for a colored black hole
of radius $r=0.55$.
}
\label{fig:showevolve}
\end{figure}

\begin{figure}
\epsfxsize=15cm
\centerline{\epsffile{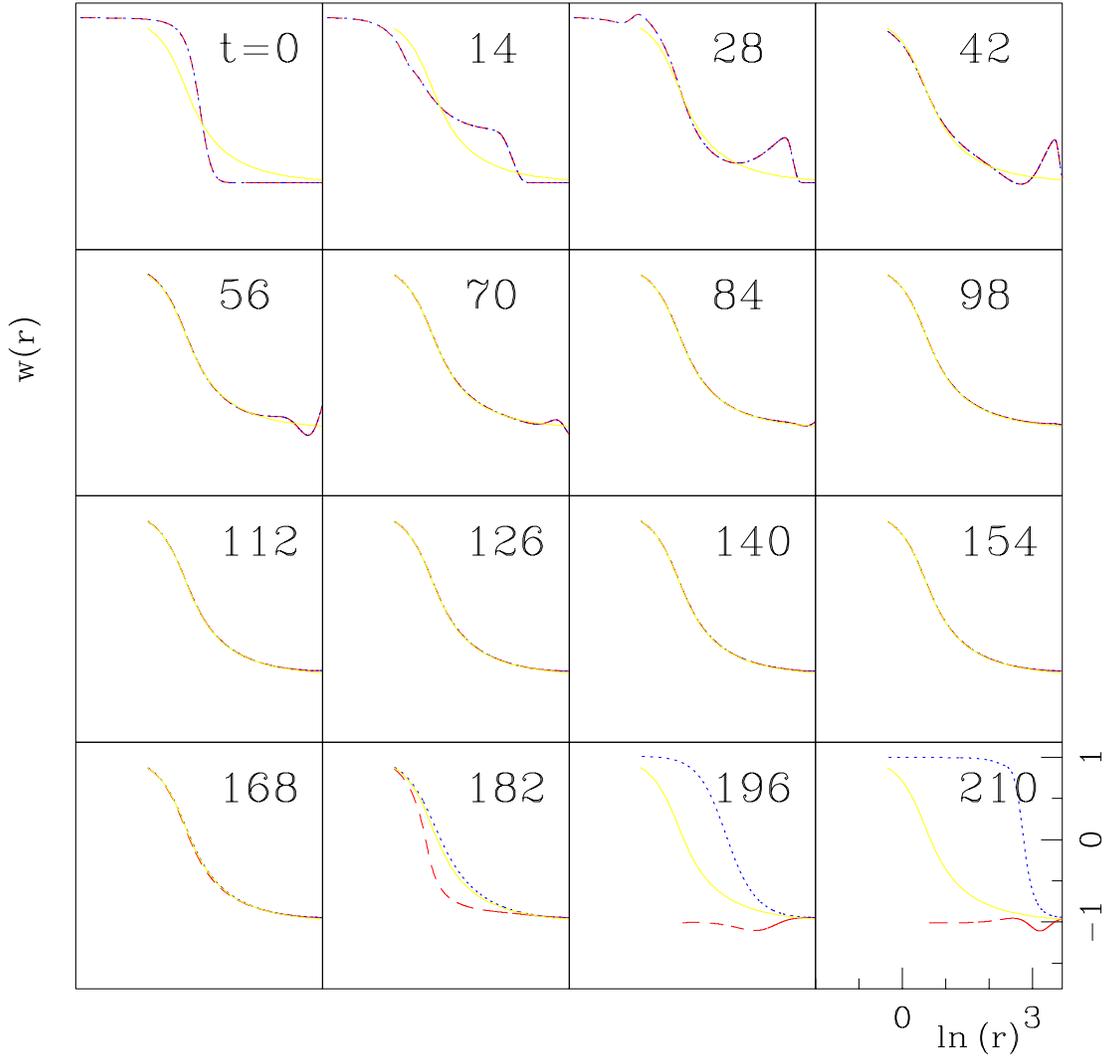}}
\caption[Snapshots of near critical evolution in logarithmic coordinates.]
{Time sequence of the Yang-Mills potential as shown
in Figure~\ref{fig:showevolve}---here the waveforms are plotted using
a logarithmic radial coordinate.
}
\label{fig:showevolve_log}
\end{figure}

It quickly becomes clear that these regions are well defined in
parameter space and it becomes natural to try and find
the boundary separating them.  Indeed, the problem reduces to tuning
a single parameter across the threshold between these two types of collapse
dynamics.  What we find is that at the threshold between these two types of
dynamics a static solution emerges
as the intermediate attractor for this new
near-critical collapse.  The Type~II side of this threshold exhibits the
following near-critical evolution.
As the field collapses, a finite mass black hole is formed.  The remaining
field exterior to the black hole then approaches a static, non-trivial
configuration.
After some time the field then departs from the static configuration and
disperses to infinity leaving the original black hole virtually unchanged.
The gauge potential in this case approaches the
vacuum value of
$w = +1$ and thus this side of the threshold is part of the generalized
Type~II
collapse dynamics.
The Type~I side of this threshold has, of course, a very similar
development initially.  Again, a black hole of
finite mass is formed exactly as before and is followed by the approach
of the exterior Yang-Mills field
to the static solution.  However, as the field leaves
the static solution, most of the matter collapses into the black hole with only
a small portion radiating off to infinity.  The gauge potential, in this case, 
approaches the vacuum state of $w=-1$.  Figures~\ref{fig:showevolve}
and~\ref{fig:showevolve_log} show the development of these two cases overlayed 
with the static, intermediate attractor which is the exactly critical
solution.

Since we are in the super-critical regime, these new, exactly critical
solutions are themselves
black holes with Yang-Mills hair.  Indeed, these attractors
are the colored black holes discovered by Bizon and independently by
Volkov and Galtsov \cite{Bizon,VolGal} shortly
after the discovery of the Bartnik-Mckinnon solutions.  In general, these
colored, static black hole solutions are described
by their horizon radius
(or, relatedly, their mass) $r_h$ and the number $n$ of zero
crossings of the Yang-Mills potential $w$.  The solutions which serve here as
intermediate attractors in the dynamical collapse
are the lowest lying colored black holes with $n=1$.  Subsequent to their
discovery, it was shown that these black holes are unstable.  In particular,
the colored black holes with label $n$ have $2n$ unstable 
modes \cite{StraumannZhou,Zhou,Volkovetal}.  Of
these $2n$ unstable modes, $n$ arise from spherically symmetric
perturbations in the gravitational sector
away from staticity and another $n$ unstable modes arise from perturbations
in the sphaleron sector, {\it i.e.} away from the magnetic {\em ansatz} for the
Yang-Mills fields.  Since we are solving the nonlinear evolution equations
while remaining within the magnetic {\em ansatz} we would expect our data to excite 
all the unstable modes within the gravitational sector while exciting
none of the unstable modes in the sphaleron sector.  As a result, we
would expect the $n=1$ family of solutions to have a single unstable
mode, thus making it relevant to critical collapse.
The picture of this new
critical solution
in the black hole regime having a single unstable mode thus
accords perfectly with
previous understanding of both Type~I and Type~II collapse.

This new critical behavior within the black hole regime is, in many ways,
qualitatively similar to that of the Type~I collapse.  Solutions 
in one of these new interpolating families asymptote to one of the static, 
colored black hole solutions,
$Y_1(r;r_h)$.
Initial data for collapsing configurations
close to one of the critical solutions will approach 
$Y_1(r;r_h)$, and remain in its
vicinity for some amount of time $T$, as measured, for example,
by an observer at infinity.
The evolution will then peel off the static solution with the remaining
field either dispersing to infinity or collapsing and adding a finite
amount of mass to the already present black hole.  One can quantify
the amount of time spent near the static solution in the same way as
with the Type~I collapse.  The
time, as measured by an asymptotic observer,
is $T \approx -\lambda \ln |p-p^{*}|$, where the coefficient $\lambda$
is the characteristic time scale for the collapse of the unstable solution
$Y_1(r;r_h)$ or, in other words, the inverse Lyapounov exponent of the 
single unstable mode.

\begin{figure}
\epsfxsize=7.5cm
\centerline{\epsffile{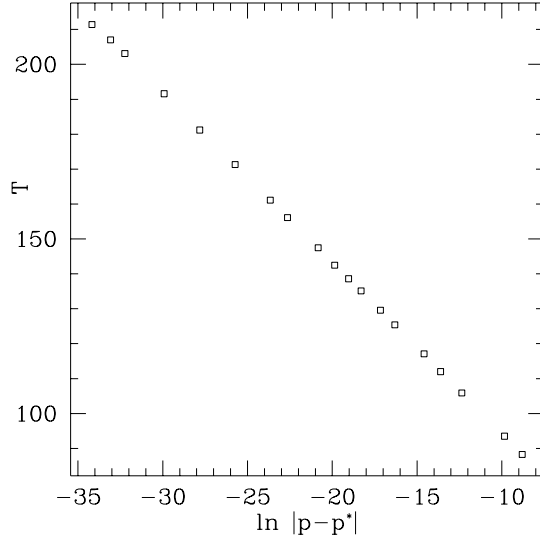}}
\caption[Scaling relation for the proper time of an asymptotic observer.]
{Plot of the elapsed time (as measured by an observer at infinity)
in generalized Type~II collapse before the zero of the $w$ field crosses $r=35$.
As the exactly critical solution is approached in parameter space, {\it
i.e.} $p\rightarrow p^{*}$, the dynamical fields spend more time
on the critical solution; specifically, the lifetime of the 
static configuration is  $T \approx -\lambda \ln |p-p^{*}|$.
Since the early part of the dynamics is unchanged as $p\to p^\star$,
the total time before the lingering field
escapes and passes a finite radial value depends only on
$|p-p^{*}|$.  Thus we can estimate $\lambda$ by measuring this total time
as a function of $\ln |p-p^{*}|$.  A straightforward least squares fit of
this data yields $\lambda\approx 4.88$.
}
\label{fig:time_exponent}
\end{figure}

With our evolutions, we confirm this relation and can calculate the value
of $\lambda$ in the manner described in \cite{ChopBizChmj}.  For
Figures~\ref{fig:showevolve} and \ref{fig:showevolve_log}, where the static 
solution has a black hole radius of $r_h \approx 0.55$, we confirm the
linear relation and find $\lambda\approx 4.88$ as shown in
Figure~\ref{fig:time_exponent}.

\begin{figure}
\epsfxsize=7.5cm
\centerline{\epsffile{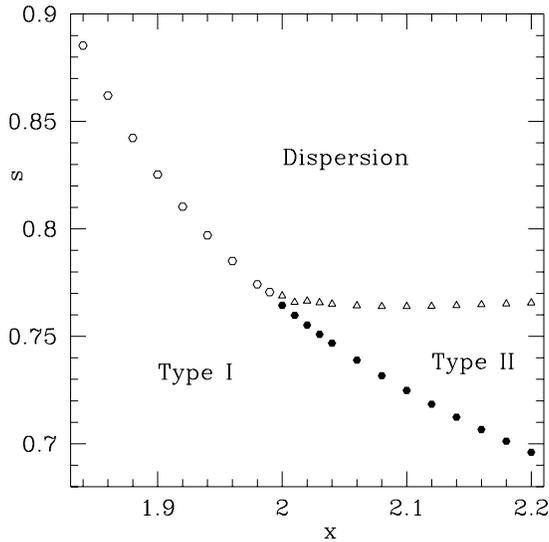}}
\caption[Phase space for kink data.]
{Plot of the phase space for the two parameter family of
kink data given by~(\ref{eq:kink}).  
Each point represents a critical solution at a
level of $|p-p^{*}|/p^{*} \ltwid 10^{-4}$ and
sits on one of the thresholds separating two of the three phases.
The open hexagons represent Type~I critical solutions which
separate configurations that disperse from those that form a finite
size black hole.  We confirm that the intermediate attractor in this
case is the $n=1$
Bartnik-Mckinnon solution, as found in \cite{ChopBizChmj}.  The open triangles 
represent Type~II critical solutions which separate configurations
that disperse from those that form an infinitesimal black hole.  The
filled hexagons represent the new black hole critical solutions
separating the generalized Type~I and Type~II collapse dynamics as described
in the text.  These critical solutions are the $n=1$ colored black
holes.  They are parameterized by their horizon radius, such that as the
``triple point'' is approached along the critical line, the size of
the black holes goes to zero.
}
\label{fig:phase_space}
\end{figure}

Once we have established the existence of these new, colored critical
solutions, it then becomes straightforward to map out parameter space
for our choice of initial data.  Figure~\ref{fig:phase_space} does this for
the main region of interest.

Figure~\ref{fig:phase_space} should be compared to Figure 1 in
\cite{ChopBizChmj}.  There, it
was suggested that there was a coexistence between black holes exhibiting
both types of critical behavior.  However, that suggestion was based 
on evolutions which 
did not actually continue to the future of the resulting black hole.  Thus, up
to this point, only the region marked ``dispersion" had been well
explored.  Evolutions had been performed in the black hole regime, but
detailed understanding of what happened in collapse was largely 
limited to times prior
to the formation of the black hole.  Now, with our use of apparent horizon
boundary conditions, we can explore more completely this super-critical
region.  Thus, what is actually present is two well-demarcated regions
in the black hole regime separated by a critical line, as described above.

\section{Discussion}
\label{sec:discussion}

We have presented evidence for the existence of a new type of
critical phenomena within the black hole regime of the spherically
symmetric Einstein-Yang-Mills model.  Previous
work on critical behavior in this model has established that
both Type~I and Type~II critical solutions emerge in sufficiently
general families of initial data.  However, that earlier work was
unable to accurately evolve to the future of black hole formation.  Using a
different coordinate system and horizon excision techniques, we
are able to evolve well into the black hole regime.  We find that
these two different types of critical collapse can be generalized to types
of collapsing spacetimes with a distinguishing characteristic
being the asymptotic vacuum value of the Yang-Mills potential.
These two ``phases" of black hole formation are separated by a critical line
in phase space such that dynamically forming black hole solutions
exactly at criticality
are the static, $n=1$ colored black holes \cite{Bizon,VolGal}.
This, of course, fits nicely within our conceptual framework that
critical solutions are intermediate attractors of co-dimension one.

It is worth emphasizing a few observations at this point.  
First, it is important  
to note that the mass of the final Schwarzschild black hole in the evolution 
will be 
discontinuous across this new critical line in Figure~\ref{fig:phase_space}.  
(In fact, indications of jumps in the black hole mass-spectrum in the 
super-critical regime were present in earlier 
calculations~\cite{BizonChmajPrivate}.)
Naively, this might be somewhat surprising since it
means that nearby
configurations in the space of initial data can have, in principle, very  
different sized black holes as their end-states.  

Further, we note that when the colored black holes were originally 
investigated, it was found that the $n$th static colored black hole 
is parameterized by its horizon radius, $r_h$.  In addition, as   
$r_h \rightarrow 0$, the black hole solution approaches 
the corresponding $n$th  
Bartnik-Mckinnon solution.  In the dynamical context discussed here 
for the $n=1$ solutions, one notices an interesting manifestation of 
this result.  The colored black hole 
attractors separating the Type~I and Type~II regions
in Figure~\ref{fig:phase_space}, are likewise parameterized by their 
horizon radius in such a way that as the ``triple point" is approached, 
$r_h \rightarrow 0$, and the Bartnik-Mckinnon solution emerges as the 
relevant attractor separating Type~I collapse from dispersion.

Finally, it is natural to ask if other collapsing systems might exhibit
similar behavior.  Indeed, one can simply look in the literature on
static, hairy black holes and point to a number of systems that one
would conjecture should have a ``phase diagram" similar to
Figure~\ref{fig:phase_space}.  Such theories should include
Einstein-Yang-Mills coupled to a dilaton (EYMD) and the
Einstein-Skyrme system.  However, there are
other examples of systems which exhibit both Type~I and Type~II critical
behavior.  These include a massive scalar field and (it is
conjectured) a massive, charged scalar field \cite{BradChamGonc}.
These latter systems are
particularly intriguing in light of the results presented here
because though they do exhibit Type~I
and Type~II critical
behavior, no-hair theorems
suggest that these systems do not have static
black hole solutions.  It should thus be of interest to 
investigate the nature of the
super-critical regime for these, and possibly other, systems.

\section*{Acknowledgments}
\label{sec:ack}
This research has been supported in part by NSF grants 
PHY93-18152,  PHY94-07194 and PHY97-22068.  MWC would like 
to thank P.~Bizon and T.~Chmaj for useful 
communications~\cite{BizonChmajPrivate}, 
including descriptions of unpublished results which indicated that the
parameter-space regime studied here might lead to colored black holes. 

\appendix
\section*{}
\label{sec:app}
We collect here some of the details involved in deriving the equations
of motion used in the paper, and to present a somewhat
more general framework for
future work on the Einstein-Yang-Mills system (EYM).
The action for our model is
\be
S = \int d^4x \sqrt{-g} \left[{R \over 16\pi G}
     - {1\over g^2} F^{a}_{\mu\nu} F^{a\mu\nu}
     \right]
\ee
where $F^{a}_{\mu\nu}$ is the Yang-Mills field strength tensor.
As usual, Greek indices range over the four spacetime dimensions and
Latin indices indicate group indices.  Spacetime
indices are raised and lowered by the metric $g_{\mu\nu}$ while we
will not lower group indices.  Thus, repeated upper group (Latin) indices will 
be summed over.

The equations of motion are found by varying the action with respect
to the fields.  Varying with respect to the metric yields the Einstein
equations
\beq
{1\over 16\pi G} G_{\mu\nu} & = & T_{\mu\nu}   \nonumber\\
& = & {1 \over g^2}(2 F^{a}_{\mu\lambda} F^a{}_\nu{}^\lambda
    - {1\over2}
     g_{\mu\nu} F^{a}_{\alpha\beta} F^{a\alpha\beta} ) .
\eeq
Varying with respect to the connection $A^{a}_{\mu}$ gives
\be
D_{\mu} F^{a\mu\nu} = \nabla_{\mu} F^{a\mu\nu} + \epsilon^{abc} A_{\mu}^{b} 
F^{c\mu\nu} = 0 .
\ee

With the general equations in hand we are in a position to make
some simplifying assumptions.  To begin, we assume that we have an $SU(2)$
gauge field.
In addition we will assume spherical symmetry.
The most general parameterization of the spacetime
metric is then
\be
ds^2 = (-\alpha^2 + a^2 \beta^2) dt^2 + 2 a^2\beta dt dr + a^2 dr^2 +
      b^2 r^2 d\Omega^2
\ee
where the metric coefficients $\alpha, \beta, a$ and $b$
will depend on the temporal and radial
coordinates
$t$ and $r$ and $d\Omega^2$ is the usual metric on the unit
two-sphere.
Likewise, the most general parameterization of the gauge connection
is now
$$
A = u\,\tau^r dt + v\,\tau^r dr
   + (w\,\tau^{\theta} + \tilde{w}\,\tau^{\phi} )d\theta
   + (\cot\theta \,\tau^r + w\,\tau^{\phi}
   - \tilde{w} \,\tau^{\theta} ) \sin\theta d\phi.
$$
where the coefficients $u, v, w$ and $\tilde{w}$ will all depend on
$t$ and $r$.  The $\tau^a$ are the spherical projection of the Pauli
spin matrices and form an anti-Hermitian basis for the group $SU(2)$
satisfying $[\tau^{a},\tau^{b}] = \epsilon^{abc}\tau^{c}$ with $a,b,c
\in \{r,\theta,\phi\}$.
We note that this connection is
invariant under a gauge transformation of the form
$U = e^{\psi(t,r) \tau^{r}}$.  The field strength derived from
this connection is
\beq
F & = & \tau^{r} (\dot{v} - u') dt\wedge dr  \nonumber\\
  & & + \left[ (\dot{w}- u\tilde{w})dt + (w'- v\tilde{w}) dr \right]
       \wedge(\tau^{\theta} d\theta +
     \tau^{\phi} \sin\theta d\phi)    \nonumber\\
  & & + \left[ (\dot{\tilde{w}}+ uw)dt + (\tilde{w}'+ vw) dr \right]
       \wedge(\tau^{\phi} d\theta -
     \tau^{\theta} \sin\theta d\phi)  \nonumber\\
  & & - (1 - w^2 -\tilde{w}^2) \tau^{r}
    d\theta\wedge \sin\theta d\phi
\eeq

With these assumptions,
the equations of motion for the Yang-Mills fields 
can now be written in first order (in time) form as
\beq
\label{eom_first}
\dot{\Pi} & = & \left[ \beta\Pi + {\alpha \over a} \Phi \right]' + uP
     - v \left( \beta P + {\alpha\over a} Q \right) +
     {\alpha a\over b^2 r^2} w (1-w^2-\tilde{w}^2)
      \\
\dot{P}   & = & \left[ \beta P + {\alpha \over a} Q \right]' - u\Pi
     + v \left( \beta\Pi + {\alpha \over a} \Phi \right) +
     {\alpha a\over b^2 r^2} \tilde{w} (1-w^2-\tilde{w}^2)
\eeq
\beq
Y' & = &
      \tilde{w}\Pi - wP  \\
\dot{Y} & = &
      {\alpha \over a}(\tilde{w}\Phi - wQ) + \beta(\tilde{w}\Pi
       - wP )
\eeq
where we have used the definitions
\beq
\label{eq:defPi}
\Pi & = & {a \over \alpha} \left[ \dot{w} - u\tilde{w} -
            \beta(w' - v\tilde{w}) \right] \\
\label{eq:defPhi}
\Phi & = & w' - v\tilde{w}   \\
P & = & {a \over \alpha} \left[ \dot{\tilde{w}} + uw -
            \beta(\tilde{w}' + vw) \right] \\
\label{eq:defQ}
Q & = & \tilde{w}' + vw   \\
\label{eq:defY}
Y & = & {b^2 r^2 \over 2\alpha a} (\dot{v} - u')  \, .
\eeq
We also have evolution equations for $\Phi$ and $Q$:
\beq
\dot{\Phi} & = & \left[ {\alpha \over a} \Pi + \beta\Phi \right]' + uQ
        - v \left[{\alpha\over a}P + \beta Q \right]
        - \tilde{w} \, { 2 \alpha a \over b^2 r^2 } Y   \\
\dot{Q}    & = & \left[ {\alpha \over a} P + \beta Q \right]'     - u\Phi
        + v \left[ {\alpha\over a}\Pi + \beta \Phi \right]
        + w \, { 2 \alpha a \over b^2 r^2 } Y    \, ,
\eeq
which follow from differentiation of~(\ref{eq:defPhi}) and (\ref{eq:defQ})
with respect to time, and elimination of explicit time derivatives using
the definitions~(\ref{eq:defPi}-\ref{eq:defY}).

Finally, the relevant Einstein equations are the evolution equations
for the metric components and the components of the extrinsic
curvature, together with the Hamiltonian and momentum constraints, as
follows:
\beq
\dot{a} & = & - a \alpha  K^r{}_r + \left( a\beta\right) '  \\
\label{eq:bdot}
\dot{b} & = & - \alpha b  K^{\theta}{}_{\theta} + {\beta \over r}\left(
rb\right)'
\eeq
\beq
\dot{K}^r{}_r & = & \beta  K^r{}_r ' + \alpha  K^r{}_r K -
  {1 \over a}\left({\alpha ' \over a}\right)'
   - {2\alpha \over arb}\left[{(rb)' \over a}\right]' + 4\pi G \alpha
  \left[ S - \rho - 2 S^r{}_r \right]   \\
\dot{K}^{\theta}{}_{\theta} & = & \beta  K^{\theta}{}_{\theta} ' +
\alpha  K^{\theta}{}_{\theta} K + {\alpha \over (rb)^2} -
{1 \over a(rb)^2} \left({\alpha r b \over a}(rb)'\right)' + 4\pi G \alpha
  \left[ S - \rho - 2 S^\theta{}_\theta \right]
\eeq
\be
-{2 \over arb}\left[\left({\left( rb\right)' \over a}\right)' +
{1 \over rb}\left(\left({rb \over a}\left(
rb\right)'\right)'-a\right)\right] + 4
K^r{}_r K^{\theta}{}_{\theta} + 2 K^{\theta}{}_{\theta}^2 =
16\pi G\rho
\ee
\be
\label{eom_last}
- {\left( rb\right)' \over rb}\left( K^{\theta}{}_{\theta} - K^r{}_r\right)-
K^{\theta}{}_{\theta}' = 4\pi G j_{r} \, .
\ee
Here, $\rho$ is the energy density, $j_r$ is the momentum density and
$S^r{}_r$, $S^\theta{}_\theta$ and $S^\phi{}_\phi$ are the stress energy
components projected onto our spacelike hypersurface. Explicitly,
we have
\be
\rho = {1\over4g^2} \left\{ {4 Y^2 \over b^4 r^4}
   + { (1-w^2-\tilde{w}^2)^2 \over b^4 r^4} + {2 \over b^2 r^2 a^2}
   \left[ Q^2 + \Phi^2 + P^2 + \Pi^2 \right] \right\}
\ee
\be
S^r{}_r = {1\over4g^2} \left\{ - {4 Y^2 \over b^4 r^4 }
    - { (1-w^2-\tilde{w}^2)^2 \over b^4 r^4} + {2 \over b^2 r^2 a^2}
   \left[ Q^2 + \Phi^2 + P^2 + \Pi^2 \right]   \right\}
\ee
\be
S^\theta{}_\theta = {1\over4g^2}
     \left\{ {4 Y^2 \over b^4 r^4}
   + { (1-w^2-\tilde{w}^2)^2 \over b^4 r^4} \right\}
\ee
\be
S^\phi{}_\phi = S^\theta{}_\theta
\ee
\be
j_{r} = - {1 \over g^2 a b^2 r^2} \left( \Pi\Phi + PQ \right) .
\ee

As can be seen, we have chosen to write things in a fairly general form.
Indeed, this is the most general EYM theory we could consider in spherical
symmetry.  We can thus
consider more general matter configurations, together with a
greater variety of coordinate systems than have been considered to date.
Subsets of these equations have been investigated before in different
contexts.  For example,
static versions of these equations have been studied in order to
understand particle-like solutions and colored black holes.
In addition,
\cite{ChopBizChmj} evolved a version of these equations in polar
($K^{\theta}{}_{\theta} = 0$), areal ($b=1$) coordinates with the additional 
assumption on the Yang-Mills field that the connection $A_\mu{}^a$ was purely 
magnetic.

For our work here, we will also make this ``magnetic {\em ansatz}."
More specifically, we will
assume that the electric charge density, $Y$, is identically zero.
In addition to this {\em ansatz},
we also make the following gauge choice:  $v=0$.  Making this gauge choice
leaves the connection invariant under a gauge transformation of the form
$U = e^{\psi(t) \tau^{r}}$.  From the definition for $Y$ in equation~(\ref{eq:defY}) 
we see that
the function $u$, by these assumptions, depends only on $t$.  However,
the residual gauge freedom implies that $u$ is arbitrary up to a function of
$t$.  We can thus completely fix the remaining gauge freedom by choosing
$u=0$ as well.
A consequence of these choices is that the remaining coefficients of
the gauge connection will no longer be independent and one of them (we
choose $\tilde{w}$)
can be set to zero without loss of generality.  Thus, in the context of this
particular problem, the magnetic {\em ansatz} has effectively set all the fields
but $w$ to zero.

As discussed in the body of the paper, we want to be able to evolve
the system for long
times to the future of any black hole formation.  To this end, we choose
a coordinate system with maximal slicing:
$K = K^r{}_r + K^\theta{}_\theta + K^\phi{}_\phi = 0$.  We will retain the
choice of areal, or radial, coordinates ($b=1$)
so that the coordinate $r$ is immediately related
to the area of origin-centered two-spheres.

Given all these assumptions, the equations which we must solve simplify
considerably.
The evolution equations are
\beq
\dot{\Pi} & = & \left[ \beta\Pi + {\alpha \over a} \Phi \right]'
     + {\alpha a\over r^2} w (1-w^2)  \\
\dot{\Phi} & = & \left[ {\alpha \over a} \Pi + \beta\Phi \right]'
          \\
\dot{w}         & = & {\alpha \over a} \Pi + \beta w'  
\eeq
and the constraint equations are
\beq
w' & = & \Phi  \\
\alpha'' & = &  \alpha' \left( {a'\over a} - {2\over r} \right)
         + {2\alpha\over r^2} \left( a^2 - 1 + {2r a'\over a} \right)
 + 4\pi G \alpha \left( S - 3 \rho \right)   \\
a' & = & a {1-a^2 \over 2r} + {3\over2} ra^3 K_{\theta}{}^{\theta}{}^2
 + 4\pi G r a \rho   \\
K_{\theta}{}^{\theta}{}' & = & -{3\over r} K_{\theta}{}^{\theta}
    + 4\pi G \left( {\Pi\Phi\over g^2 a r^2}  \right)
\eeq
where the matter stress-energy terms are given by
\beq
S - 3 \rho & \, = & \,
      { a^2 (1-w^2)^2 \over 2 g^2 r^4} + {1 \over g^2 r^2 }
         \left( \Phi^2 + \Pi^2 \right)      \\
\rho & \, = & \,
      { a^2 (1-w^2)^2 \over 4 g^2 r^4} + {1 \over 2 g^2 r^2 }
         \left( \Phi^2 + \Pi^2 \right) .
\eeq
Finally, we have an algebraic condition for the non-vanishing component
of the shift vector
\beq
\beta = \alpha r K_{\theta}{}^{\theta} \, ,
\eeq
which follows from~(\ref{eq:bdot}) with $b(r,t) \equiv 1$.


\end{document}